\documentstyle[aps,twocolumn,prl]{revtex}

\begin{document}
% \draft

%%% K. Hirota %%%
\input BoxedEPS
\SetTexturesEPSFSpecial
\HideDisplacementBoxes
%\input EPSF.sty
%%% K. Hirota %%%

\title{Spin Dynamical Properties of the Layered Perovskite
La$_{1.2}$Sr$_{1.8}$Mn$_{2}$O$_{7}$}
\author{H. Fujioka,$^{\dag}$ M. Kubota,$^{\ddag}$ K. Hirota,$^{\dag}$ H.
Yoshizawa,$^{\ddag}$ Y. Moritomo$^{\P}$ and Y. Endoh,$^{\dag}$}
\address{$^{\dag}$CREST, Department of Physics, Tohoku University, Sendai
980-8578, Japan}
\address{$^{\ddag}$Neutron Scattering Laboratory, I.S.S.P., University of Tokyo,
Tokai, Ibaraki, 319-1106, Japan}
\address{$^{\P}$Center for Integrated Research in Science and Engineering, 
Nagoya University, Nagoya, 464-01, Japan}

\date{September 23, 1998}

\twocolumn[\hsize\textwidth\columnwidth\hsize\csname @twocolumnfalse\endcsname
\maketitle

\begin{abstract}
Inelastic neutron-scattering measurements were performed on a
single crystal of the layered colossal magnetoresistance (CMR) material
La$_{1.2}$Sr$_{1.8}$Mn$_{2}$O$_{7}$ ($T_{C}\sim 120$~K).  We found that the spin
wave dispersion is almost perfectly two-dimensional with the in-plane spin
stiffness constant $D \sim 151$~meV\AA. The value is similar to that of similarly
doped La$_{1-x}$Sr$_{x}$MnO$_{3}$ though its $T_{C}$ is three times higher,
indicating a large renormalization due to low dimensionality.  There exist two
branches due to a coupling between layers {\em within} a double-layer. The out-of-plane coupling is
about 30\% of the in-plane coupling though the Mn-O bond lengths are similar. 

\vspace{0.5cm}
\vspace{0.5cm}

\end{abstract}
]

%\vskip .5in
%\pacs{}

%\newpage
\section{Introduction}

The layered perovskite Mn oxide La$_{2-2x}$Sr$_{1+2x}$Mn$_{2}$O$_{7}$
(LSMO327), in which MnO$_{2}$ double layers and (La,Sr)$_{2}$O$_{2}$ blocking
layers are stacked alternatively, attracts much attention as
another class of colossal magnetoresistance (CMR) system. Possibly due to the
reduced dimensionality, this system exhibits an extremely large MR
at the hole concentration $x=0.4$\cite{Y.Moritomo_96} and a tunneling 
MR phenomenon at $x=0.3$\cite{T.Kimura_96} around $T_{C}$.  
Neuron-scattering study on LSMO327 single crystals ($x=0.40-0.48$) by Hirota
{\it et al.}\cite{K.Hirota_98} has revealed that the low-temperature magnetic
structure consists of planar ferromagnetic (FM) and A-type antiferromagnetic
(AFM) components, indicating a canted AF structure, where the
canting angle between neighboring planes changes from
6.3$^{\circ}$ at $x=0.40$ (nearly planner FM) to 180$^{\circ}$ at $x=0.48$ 
(A-type AF).\cite{K.Hirota_98}  The existence of the canted AFM structure is
consistent with previous studies focusing upon the structural
properties.\cite{J.F.Mitchell_96,D.N.Argyriou_97} Kubota {\it et
al.}\cite{M.Kubota_98} carried out a comprehensive powder neutron-scattering work
and established the magnetic phase diagram for $x=0.30-0.50$; there is a planar
FM phase between $x=0.32$ and $0.38$, which is smoothly connected to the canted
AFM region.  To understand the magnetic properties of LSMO327 in more detail, it
is necessary to study the excitation spectra, from which one can determine
magnetic interaction strengths and spin-spin correlation lengths.

\section{Theoretical Model}

\begin{figure} 
%%% K. Hirota %%%
\begin{center}
\BoxedEPSF{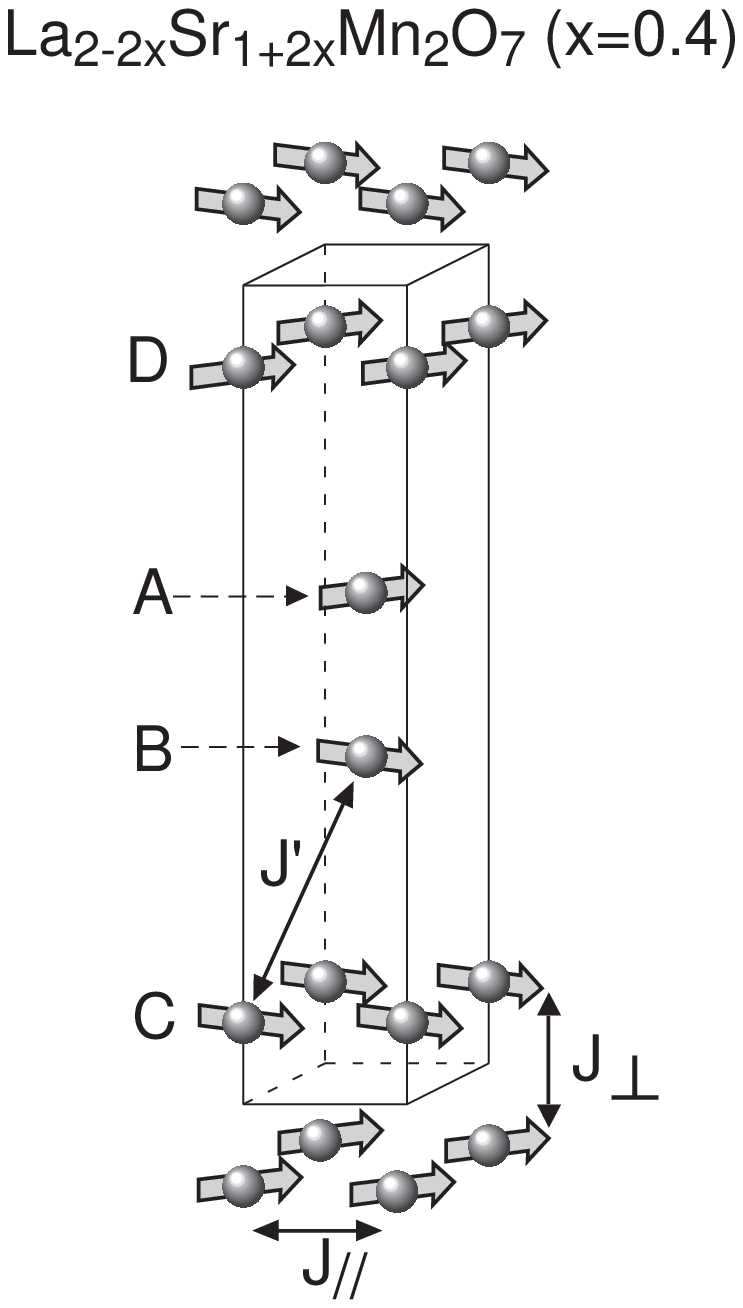 scaled 600}
\vspace{0.5cm}
\end{center}
%%% K. Hirota %%%
\caption{The magnetic spin arrangement on Mn ions in the $I4/mmm$ tetragonal
cell of La$_{1.2}$Sr$_{1.8}$Mn$_{2}$O$_{7}$. The lattice parameters are
$a=b=3.87$ and $c=20.1$~\AA~at 10~K.\protect\cite{K.Hirota_98}}
\label{Fig:Structure}
\end{figure}

Figure~\ref{Fig:Structure} shows the magnetic spin arrangement on Mn ions in the
tetragonal $I4/mmm$ cell of La$_{1.2}$Sr$_{1.8}$Mn$_{2}$O$_{7}$.  Although there
is a small canting between neighboring layers within a double-layer at
$x=0.40$, we assume a simple planar ferromagnet, which is
sufficient to consider the magnetic interactions between $x= \leq 0.40$. 
We expect that the dominant spin-spin interactions should occur between
nearest-neighbor (NN) Mn atoms, though the in-plane interaction $J_{\parallel}$
and the {\em intra}-bilayer interaction $J_{\perp}$ might be different. 
Although there is no super-exchange coupling between layers belonging to
different double-layers, there will be the {\em inter}-bilayer interaction
$J'$ through a direct exchange.  However, it is supposed to be much weaker than
$J_{\parallel}$ and $J_{\perp}$, thus we neglect it in our simple model
calculation.

Let $l=A,B,C,D$ label the four different layers as indicated in
Fig.~\ref{Fig:Structure}.  The spin Hamiltonian can then be written in the
Heisenberg form as
%%%
\begin{eqnarray} {\cal H}  & = & \frac{1}{2}\sum_{i}\sum_{l}\sum_{\delta}
J_{il\delta}{\bf S}_{i}^{l}\cdot {\bf S}_{\delta}^{l} \nonumber \\ & = &
\frac{1}{2}\sum_{i}\sum_{\delta}\left[ {\bf S}_{i}^{A}\left\{J_{\parallel}{\bf
S}_{\delta}^{A}+J_{\perp}{\bf S}_{\delta}^{B}\right\}+ {\bf
S}_{i}^{B}\left\{J_{\parallel}{\bf S}_{\delta}^{B}+J_{\perp}{\bf
S}_{\delta}^{A}\right\}\right.\nonumber \\ & &\left. +{\bf
S}_{i}^{C}\left\{J_{\parallel}{\bf S}_{\delta}^{C}+J_{\perp}{\bf
S}_{\delta}^{D}\right\} +{\bf S}_{i}^{D}\left\{J_{\parallel}{\bf
S}_{\delta}^{D}+J_{\perp}{\bf S}_{\delta}^{C}\right\}
\right],
\label{eq:Hamiltonian}
\end{eqnarray}
%%%
where $i$ denotes a unit cell and $\delta$ indicates NN sites corresponding
to a particular interaction, $J_{\parallel}$ or $J_{\perp}$.  Following the
standard approach, we make the Holstein-Primakoff
transformation\cite{T.Holstein_40} to boson creation and annihilation opperators
$a_{n}^{\dagger},a_{n}, b_{n}^{\dagger},b_{n}, c_{n}^{\dagger},c_{n},
d_{n}^{\dagger},d_{n}$, which correspond to $A, B, C, D$ layers.  By
Fourier transforming to reciprocal space and performing diagonalization, we
obtain the following dispersion relations as the eigenvalues:
%%%
\begin{eqnarray}
\hbar\omega ({\bf q}) & = & -2J_{\parallel}S\left(2-\cos aq_{x}-\cos
aq_{y}\right) \nonumber \\
& & - J_{\perp}S\left\{1 \mp |\exp\left(-i2zcq_{z}\right)|\right\}
\nonumber \\
& = & -2J_{\parallel}S\left(2-\cos aq_{x}-\cos aq_{y}\right) 
- J_{\perp}S\left(1 \mp 1\right),
\label{eq:Dispersion}
\end{eqnarray}
%%%
where $a$ and $c$ are the lattice constants and $2zc$ is the
distance between layers within a double-layer.  Although there should be four
different modes, these are classified to two modes, i.e, acoustic (A) and
optical (O), when the inter-bilayer coupling $J'$ is neglected.  Note that both
$J_{\parallel}$ and $J_{\perp}$ are negative because they are FM interactions.

By using a unitary matrix diagonalizing the Hamiltonian
Eq.~\ref{eq:Hamiltonian},  we obtain the differential scattering cross section
for spin waves in LSMO327 with $x=0.40$
%%%
\begin{eqnarray}
\frac{d^{2}\sigma}{d\Omega dE_{f}} & = &
\left(\frac{\gamma e^{2}}{mc^{2}}\right)^{2} \frac{k_{f}}{k_{i}}
\left\{\frac{1}{2} g f(Q)\right\}^{2} \left(1+\hat{Q}_{x}^{2}\right) e^{-2W(Q)}
\nonumber \\
& \times & 
\frac{S}{2} \frac{(2\pi)^{3}}{v_{0}}\frac{4}{N} \sum_{m}\sum_{\bf qG}
\left(n_{q}^{(m)}+\frac{1}{2}\pm\frac{1}{2}\right)
\\
& \times & 
\delta (\hbar\omega_{q}^{(m)}\mp\hbar\omega_{q})
\delta ({\bf Q}\mp{\bf q}-{\bf G}) \left\{1\pm\cos(2zc\cdot Q_{z})\right\}
\nonumber
\protect\label{eq:CrossSection}
\end{eqnarray}
%%%
where $\gamma$ is the gyromagnetic ratio of the neutron, $f(Q)$ is the
magnetic form factor for a Mn ion, $\exp[-2W(Q)]$ is a Debye-Waller factor,
$\hat{Q}_{x}=Q_{x}/Q$, $n_{q}^{(m)}$ is the bose factor and $m$ denotes a
mode.  Since $2zc$ is very close to $a \approx c/5$, the A-branch has
maximum intensity at $l=5n$ ($n$:integer), while the phase of O-branch is shifted
by $\pi$.

%\newpage
\section{Experimental}

La$_{1.2}$Sr$_{1.8}$Mn$_{2}$O$_{7}$ powder was prepared
by solid-state reaction using prescribed amounts of pre-dried
La$_{2}$O$_{3}$ (99.9\%), Mn$_{3}$O$_{4}$ (99.9\%), and SrCO$_{3}$ (99.99\%).
The powder mixture was calcined in the air for 3 days at 1250$^{\circ}$C 
--1400$^{\circ}$C with frequent grindings. The calcined powder was then pressed
into a rod and heated at 1450$^{\circ}$C for 24~h. Single crystals were
melt-grown in flowing 100\% O$_{2}$ in a floating zone optical image furnace
with a travelling speed of 15 mm/h. To check the sample homogeneity, we
powderized a part of single crystal and performed x-ray diffraction, which
shows no indication of impurities.

Neutron-scattering measurements were carried out on the triple-axis spectrometer
TOPAN located in the JRR-3M reactor of the Japan Atomic Energy Research
Institute (JAERI). The $(0\ 0\ 2)$ reflection of pyrolytic graphite (PG) was used
to monochromate and analyze the neutron beam, together with a PG filter to
eliminate higher order contamination. The spectrometer
was set up in two conditions in the standard triple-axis mode, typically with
the fixed final energy at 13.5~meV and the horizontal collimation of
B-100$'$-S-100$'$-B.   The sample was mounted in an Al can so as to give
the ($h\ 0\ l$) zone in the tetragonal
$I4/mmm$ notation.  We studied the same crystal used in
Ref.~\onlinecite{K.Hirota_98} (F-40), which is a single grain with mosaic spread
of $\sim 0.3^{\circ}$ full width at half maximum (FWHM).

\begin{figure}
%%% K. Hirota %%%
\begin{center}
\BoxedEPSF{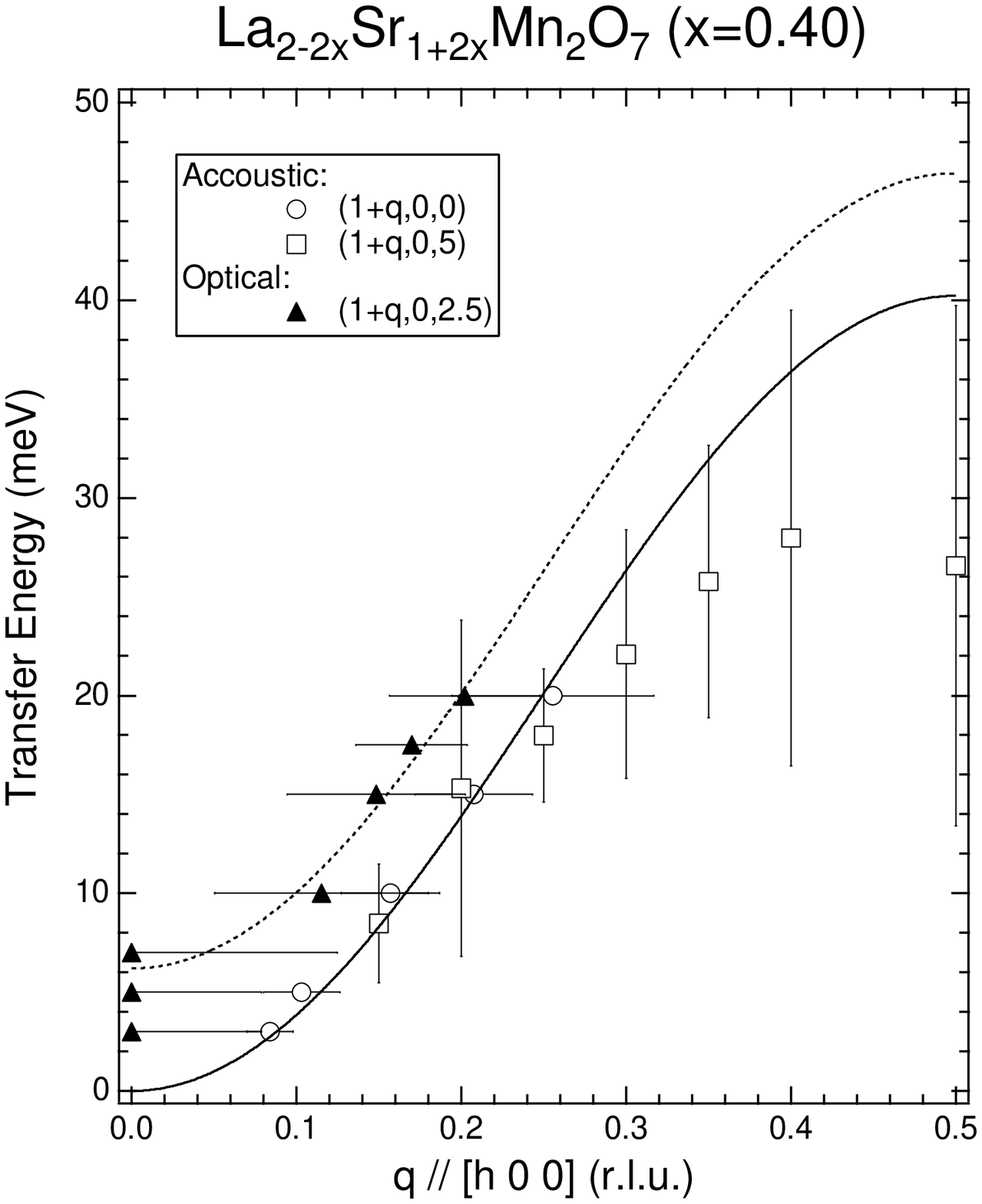 scaled 500}
\vspace{0.5cm}
\end{center}
%%% K. Hirota %%%
\caption{The dispersion relations of spin waves at 10~K.  Error bars correspond
to the FWHM of peak profiles.  Open circle and square indicate the
acoustic branch, and solid triangle indicates the opical branch. Solid and
dotted curves are obtained by fitting to Eq.~\protect\ref{eq:Dispersion} for $0 < q
\leq 0.25$~r.l.u.}
\label{Fig:Dispersion}
\end{figure}

\section{Results and Discussions}

The spin-wave dispersions along $[h\ 0\ 0]$ were measured at 10~K around $(1\ 0\
0)$ and $(1\ 0\ 5)$ for the A-branch, and $(1\ 0\ 2.5)$ for the O-branch, as
shown in Fig.~\ref{Fig:Dispersion}.  Error bars correspond to the FWHM
of peak profiles.  By fitting all the data points for $0 < q \leq 0.25$~r.l.u.\
simultaneously, we obtain $-J_{\parallel}S = 10.1$~meV and
$-J_{\perp}S=3.1$~meV.

\begin{figure}
%%% K. Hirota %%%
\begin{center}
\BoxedEPSF{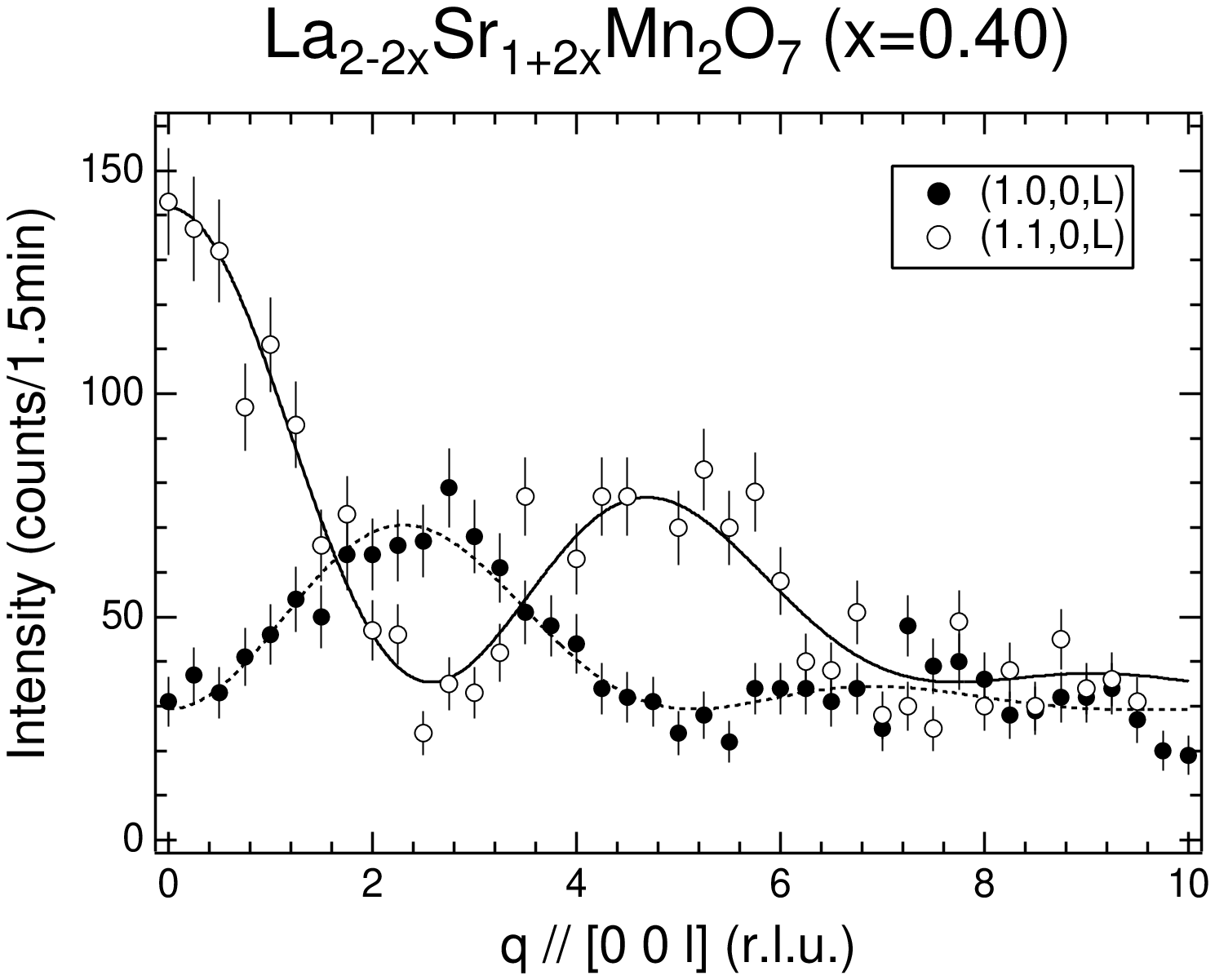 scaled 500}
\vspace{0.5cm}
\end{center}
%%% K. Hirota %%%
\caption{ The $L$-dependence of the constant-$E$ scan. The solid and open
circles indicate intensities of the acoustic and optical branch respectively.
The solid curve corresponding to the fitting with Eq.~3. The acoustic branch is
dominant at (1~0~0) and (1~0~5), and the optical branch is dominant at (1~0~2.5)
}
\label{Fig:CrossSection}
\end{figure}

To quantitatively examine the present model, we measured the $l$-dependence of
the spin-wave intensities of A and O branches at a fixed transfer energy $\Delta
E=E_{i}-E_{f}=5$~meV.  As shown in Fig.~\ref{Fig:CrossSection}, the differencial
scattering crosssection Eq.~3 is in an excellent agreement
with both the A $(1.1\ 0\ l)$ and O $(1.0\ 0\ l)$ branches.  Note that we do
not use any fitting parameters except for intensity scaling.

The results show that spin-spin correlations are significantly anisotropic. The inter-bilayer interaction is as small as we can not detect. The {\em
intra}-bilayer interaction compared with in-plane interaction, $J_{\perp}
/J_{\parallel}$ is about 0.31.  We speculate that $x^{2}-y^{2}$ orbital is
dominant in the Mn $e_{g}$ band, which enhances the double-exchange, i.e.,
ferromagnetic, interactions within a plane.  A close relation between the
magnetism and the Mn $e_{g}$ orbital degree of freedom has been also pointed out
by recent studies.\cite{K.Hirota_98,Y.Moritomo_97,Y.Moritomo_98}  The in-plane spin wave
stiffness constant $D=-J_{\parallel}Sa^2$ is about 151~meV\AA$^2$,  which is
corresponding to the nearly cubic perovsikte
La$_{1-x}$Sr$_{x}$MnO$_{3}$($x=0.2-0.3$) whose $D$ are
188~meV($x=0.3$,$T_{C}=370$~K) and 120~meV($x=0.2$, $T_{C}=310$~K)
~\cite{Hirota_97,Martin_96}.  $T_{C}$ (120~K in our material) is very reduced,
indicating a large renormalization due to low dimensionality. 

We noticed that the energy-width of constant-$Q$ scan profile becomes broad at
large $q$, particularly $q>0.25$~r.l.u.  In the same high $q$ range, the peak
position starts deviating from the dispersion curve obtained from small $q$
data using a conventional Heisenberg model Eq.~\ref{eq:Hamiltonian}.  Similar
kind of broadening and deviation are seen in other CMR systems, such as
Nd$_{0.7}$Sr$_{0.3}$MnO$_{3}$,\cite{Baca_98} which has a narrower
electronic band-width than La$_{0.7}$Sr$_{0.3}$MnO$_{3}$.  Although it is not
clear that electron-phonon coupling plays a significant role in such anomalies
in LSMO327 as suggested in Nd$_{0.7}$Sr$_{0.3}$MnO$_{3}$, it would be
interesting to study the relation between structural and magnetic properties,
particularly in their dynamics.

\section{Acknowledgments}

The authors thank S. Ishihara and S. Okamoto for constructing the theoretical model. 
This work was supported by a Grant-in-Aid for Scientific Research of MONBUSHO.

% now the references. delete or change fake bibitem. delete next three
%   lines and directly read in your .bbl file if you use bibtex.

\end{document}